\documentclass[12pt,preprint]{aastex}

\shorttitle{PRVS Pathfinder in the NIR}
\shortauthors{Ramsey et al.}

\usepackage[cmex10]{amsmath}

\newcommand{\ms}{\mbox{m s$^{-1}$}}
\newcommand{\kms}{\mbox{km s$^{-1}~$}\thinspace}
\newcommand{\mm}{$\mu$m\thinspace}
\newcommand{\eg}{{\it e.g.,}\thinspace}
\newcommand{\augapmin}{$51$\thinspace}
\newcommand{\augapmax}{$52$\thinspace}
\newcommand{\novapmin}{$50$\thinspace}
\newcommand{\novapmax}{$53$\thinspace}

\begin{document}

\title{A Pathfinder Instrument for Precision Radial Velocities in the Near-Infrared}

\author{L.W. Ramsey\altaffilmark{1}, J. Barnes\altaffilmark{2}, S.L. Redman\altaffilmark{1}, H.R.A. Jones\altaffilmark{2}, A. Wolszczan\altaffilmark{1}, S. Bongiorno\altaffilmark{1}, L. Engel\altaffilmark{1}, and J. Jenkins\altaffilmark{1}}

\altaffiltext{1}{Department of Astronomy \& Astrophysics, The Pennsylvania State University, University Park, PA 16802}
\altaffiltext{2}{Centre for Astrophysics Research, University of Hertfordshire, College Lane, Hatfield AL10 9AB, UK}

\keywords{Instrumentation: IR Spectrograph, Precision Radial Velocities}

\begin{abstract}
We have designed and tested an in-plane \'{e}chelle spectrograph configured to investigate precision radial velocities from ground-based near-infrared observations. The spectrograph operates across the spectral range of $0.9$-$1.7$~\mm\ at a spectral resolution of $R = 50,000$, and uses a liquid nitrogen-cooled  {\sc hawaii} 1K detector. Repeated measurements of the Earth's rotation via integrated Sunlight with two different instrument arrangements in the near infrared Y band have produced radial velocities with $\sim 10$~\ms\thinspace RMS over a period of several hours. The most recent instrument configuration has achieved an unbinned RMS of $7$ \ms\thinspace and suggests that infrared radial velocity precisions may be able to approach those achieved at optical wavelengths.
\end{abstract}

\begin{keywords}{Astronomical Instrumentation, Extrasolar Planets}
\end{keywords}

\section{Introduction}
The achievement of meter-per-second radial velocity precision is one of the major technological breakthroughs of recent years. Although this effort is frequently viewed as driven by the search for extra-solar planets following the discovery of the first such system in the early 1990s (Wolszczan \& Frail 1992), the quest for highly accurate radial velocity measurements was actually under development for many decades beforehand. The path from solar studies (Becker 1976; Koch \& Woehl 1984) to pioneering stellar observations using HF gas cells (Campbell \& Walker 1979; Campbell, Walker \& Yang 1988) to today's standards of long-term several \ms\thinspace (or short-term sub-\ms) RMS measurements with Keck/HIRES (e.g., Vogt et al. 1994, Butler et al, 1996), ESO~3.6m/HARPS (e.g., Pepe et al. 2000), HET/HRS (e.g., Tull 1998), VLT-UVES (Dekker et al. 2000), and AAT/UCLES (Tinney et al. 2001) is one that leads through several spectrographs, detectors, and calibration methodologies.  

To date, the only astronomical programs that have yielded sub-16~\ms\thinspace radial velocities are based on optical ($500$- to $600$-nm) \'{e}chelle spectroscopy, with velocity calibration based upon I$_2$ absorption or Th-Ar emission line references. While this wavelength region is optimal for investigating solar-like stars, it suffers when investigating cooler objects (especially in comparison with the infrared region).  Main sequence M stars are particularly inviting targets for precision radial velocity work, both because these objects represent the most numerous class of stars and their low mass makes them promising objects for detection of Earth-mass planets in the ``habitable zone" (Kasting \& Catling, 2003).  This is one of the top priorities of the Exoplanet Task Force~\footnote{http://www.nsf.gov/mps/ast/aaac/exoplanet\_task\_force/reports/aaac\_draft.pdf}.

No instruments have published radial velocity results in the near infrared with precisions approaching those in the optical, although exciting science results have been obtained in this wavelength region with hundreds of \ms\thinspace precisions. For example, Stassun, Mathieu \& Valenti (2006) characterized the orbital parameters of the brown dwarf binary \hbox{{\sc 2mass} J05352184$-$0546085} using the {\sc phoenix} spectrograph (Hinkle et al. 2003) on the Gemini South telescope. Using six nights of observations with {\sc nirspec} (McLean et al. 1998) at the Keck telescope, Martin et al. (2006) surpassed the precision achieved by previous optical radial  velocity work on late-type stars; these infrared radial velocity measurements of the nearby brown dwarf LP944-20 report an RMS precision of 360~\mbox{m s$^{-1}~$}. Blake et al. (2007) used CO bands and telluric features to reach precisions of approximately 300~\ms\thinspace around L dwarfs.

Moving beyond the CCD range to a wavelength regime where there is no tradition of precision ($< 20$ \ms) radial velocity studies is challenging because there is no legacy work to build upon. In time it should be possible to increase the sensitivity of tellurically-calibrated programs down to mean atmospheric velocities of 10-25~\ms\thinspace that have now been achieved in the optical (\eg Johnston, 2006 and Gray \& Brown, 2006).  However, more precise calibration techniques are required if infrared observations are to reach the \ms\thinspace accuracy of current state-of-the-art optical projects.

This paper describes our experiments with a laboratory-based, high-resolution infrared spectrograph configured to measure precision radial velocities. This instrument, designated the Precision Radial Velocity Spectrograph Pathfinder (hereafter, Pathfinder), was designed and constructed in the Department of Astronomy and Astrophysics at Penn State. The goal of this project was to measure the rotation velocity of the Earth by obtaining infrared spectra of integrated Sunlight to explore the limitations of making precision radial velocities in the $0.9$-$1.7$~\mm\thinspace spectral region. The most serious concern is the lack of a proven calibration technique.  NIR arrays also have issues not present in CCDs, such as inter-pixel capacitance, multiplexer cross-talk, persistence, and sensitivity to thermal background radiation.  Telluric contamination is also a major issue in the NIR, as is modal noise, since the number of modes in optical fibers decreases with wavelength.

We carried out initial observations in August 2006 and, after a series of upgrades and intermediate tests, conducted another set of observations in November 2007.  \S~2 presents the design of the instrument and the properties of its various components.  The data processing and analysis are described in \S~3.  Our conclusions, which point to the challenges that lie ahead, are given in \S~4.

\section{Spectrograph}
The science requirement of measuring radial velocities to \ms\thinspace accuracy in the infrared, with a very limited budget, required that the Pathfinder spectrograph cover the near infrared regime (no K-band sensitivity) at a spectral resolution of several tens of thousand. Pathfinder is a brass board instrument that covers the Y-, J-, and H-bands ($0.9$ to $1.7$~\mm) at a resolution of 50,000.  The instrument was assembled using existing in-house optical components that were retired or borrowed from other projects. 

The spectrograph is constructed primarily with standard optical mounting hardware on an \hbox{$8 \times 2$ foot} optical bench, but the fiber input and the detector are mounted on custom-made units. This table is located in an interior lab that has controlled access.  We did not have precision temperature monitoring hardware in place when the measurments discussed here were taken.  However, recently-installed Lakeshore PT102 platinum resistance thermometers with a Lakeshore 218 monitor indicate ambient room temperature variations are typically $0.1$ K over several hours, while spectrograph temperature variations are on the order of $0.05$ K. The spectrograph, located in an interior room in Davey Laboratory on the Penn State campus, is fed by a single fiber optic cable whose input end is located on the roof of the building in one of the rooftop observatories.

\subsection{Optical Design}

\begin{figure}
\begin{center}
\includegraphics[width=165mm,angle=-0]{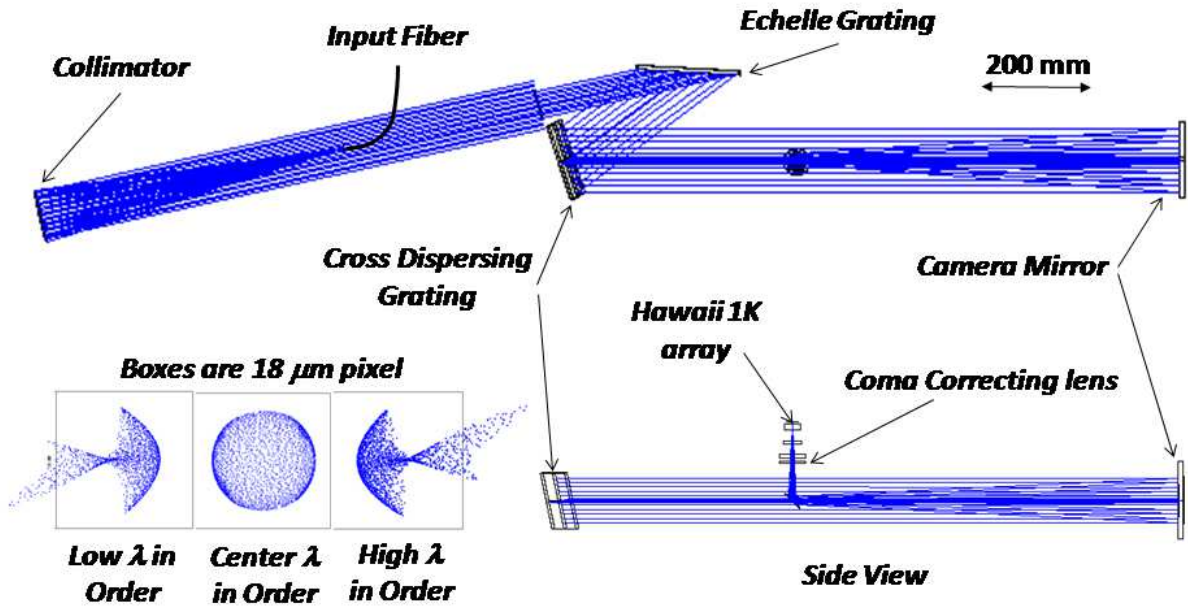}
\caption{The optical layout from a Zemax model of the pathfinder is illustrated as well as the theoretical imaging performance.  The top of this figure is a top view in the plane of the echelle dispersion.  The bottom right shows a side view from the cross disperser to the final focal plane.  The bottom left shows the expected spots at the center and edges of an order centered on the  {\sc hawaii} 1K array.  The boxes are $18~\mu m^2$ and are approximately the size of one pixel.}
\label{plotone}
\end{center}
\end{figure}

We will adopt the notation of Schroeder (2000) in our description of the Pathfinder optical design.  An optical layout of the system is displayed in Figure~\ref{plotone}.  The instrument has a straightforward in-plane configuration with a R2 \'{e}chelle and a 100~mm collimated beam diameter.  The \'{e}chelle is operated at an unusually large $\theta$ angle of 11.66$^{\circ}$, which is required to achieve the desired resolving power.  The collimator is a simple 609~mm-focal length parabola.  The collimated beam is dispersed by a gold-coated 31~line~mm$^{-1}$ \hbox{$\Theta_{\rm blaze} = 63.4^{\circ}$} R2 \'{e}chelle on a \hbox{110 $\times$ 210 mm} Zerodur substrate.  Given that \hbox{$\theta = 11.66^{\circ}$,} the incident angle, \hbox{$\alpha = \Theta_{\rm blaze} + \theta \approx 75^{\circ}$,} produces a substantial vignetting of the collimated beam by the \hbox{200 $\times$ 100 mm} ruled area of the grating. The effective collimated beam at the grating is 100~mm high and  approximately 50~mm wide.  The diffracted beam has an anamorphic magnification in the dispersion direction of $\sim 2.3$ that is cross-dispersed by a 150~lines~mm$^{-1}$, \hbox{$\Theta_{\rm blaze} = 2.15^{\circ}$} grating. The efficiency of this in-house (but far from optimal) cross disperser is low ($\sim$20\%) in the wavelength band ($0.95$-$1.35$~\mm) of our initial observations.

The Pathfinder camera is, like the collimator, a simple 154~mm diameter, 916~mm focal length parabola.  A weak lens near the focus corrects much of the coma; this lens gives the camera an effective focal length of 800~mm.   The camera beam is folded so as to allow the dewar to be mounted vertically.  This configuration allows the detector coolant (liquid nitrogen; hereafter, LN$_2$) to always cover the cold plate on which the detector is mounted.

Fig.~\ref{plotone} also shows the camera layout and spot diagram over the field of view of the detector in the focal plane.  It is clear that the effects of coma are controlled to about 1~pixel at the edges of the detector; this is of considerable importance as there are only about 2.5 detector pixels per resolution element.  The variation of the spot shape over the camera field imposes a limit on the radial velocity precision achievable in our system due to the limited number of calibration lines.  As this is essentially a variation with wavelength, fundamentally the quantity we are measuring, it can only be well compensated if there is a calibration line in close proximity.  While limited, this implementation was judged sufficient for our initial tests.

\subsection{Detector \& Dewar system}

\begin{figure}
\begin{center}
\includegraphics[width=150mm,angle=-0]{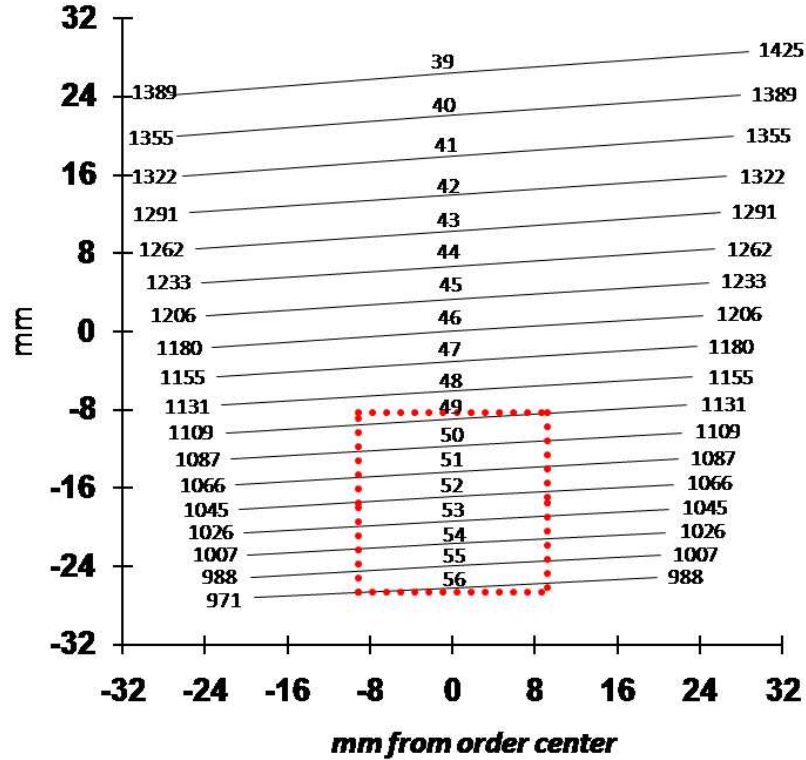}
\caption{Each line represents one free spectral range of the order number given at the center.  Beginning and ending wavelength in \mm are also labeled.  The coverage given by the array used by pathfinder for the observations discussed here is shown by the dotted box.}
\label{plottwo}
\end{center}
\end{figure}

Pathfinder's detector is a  {\sc hawaii} 1K ($1024$ pix$^2$) science-grade array provided through IR Labs in Tucson Arizona and uses a SDSU Gen II controller.  The dewar is cooled with LN$_2$.  Inside the dewar radiation shield a filter holder is mounted to the same cooling block as the detector. The temperature is continuously monitored and is typically 82.3~K. The thermal background is suppressed with 6-mm PK50 plate that attenuates radiation beyond 2~\mm\ and a short-pass filter with a cut-off at 1.8~\mm\ that has transmission less than $10^{-3}$ out to 3~\mm.  In operation the detector has a gain of 6.1~e$^-$/ADU and a read noise of 20~e$^-$.  The bias voltage has been optimized for well capacity.  We use single destructive readouts. Tests have shown that the linearity of the array is within 1\% of linear for ADU counts below $3 \times 10^{4}$, which is our operational range. The effect of pixel-to-pixel variation of quantum efficiency, which is about 6.7\% in our array, is reduced by averaging over the height of the spectrograph slit and by dividing by the flat-field spectra.  As in any precision measurement, high S/N flat fields are crucial. The optical design yields a spectroscopic resolution of 50,000 with $2.5$~pixels per resolution element ($\sim$2.5 km/s/pixel). At $1.0$~\mm\ approximately $44$\% of the Free Spectral Range (FSR) in an \'{e}chelle order is covered by the detector; this value decreases to $33$\% at $1.6$~\mm.  Figure~\ref{plottwo} illustrates the coverage of the detector in the $0.98$-$1.4$~\mm\ region.

\subsection{Fiber Input}

\begin{figure}
\begin{center}
\includegraphics[height=30mm,angle=-0]{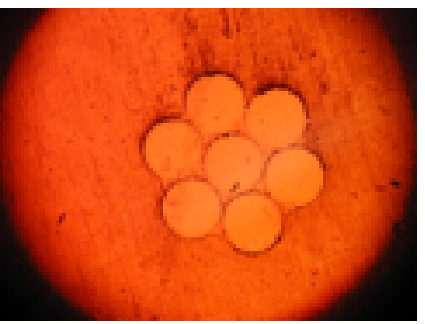}
\includegraphics[height=30mm,angle=-0]{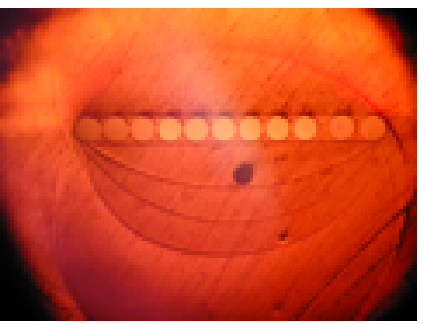}
\includegraphics[height=30mm,angle=-0]{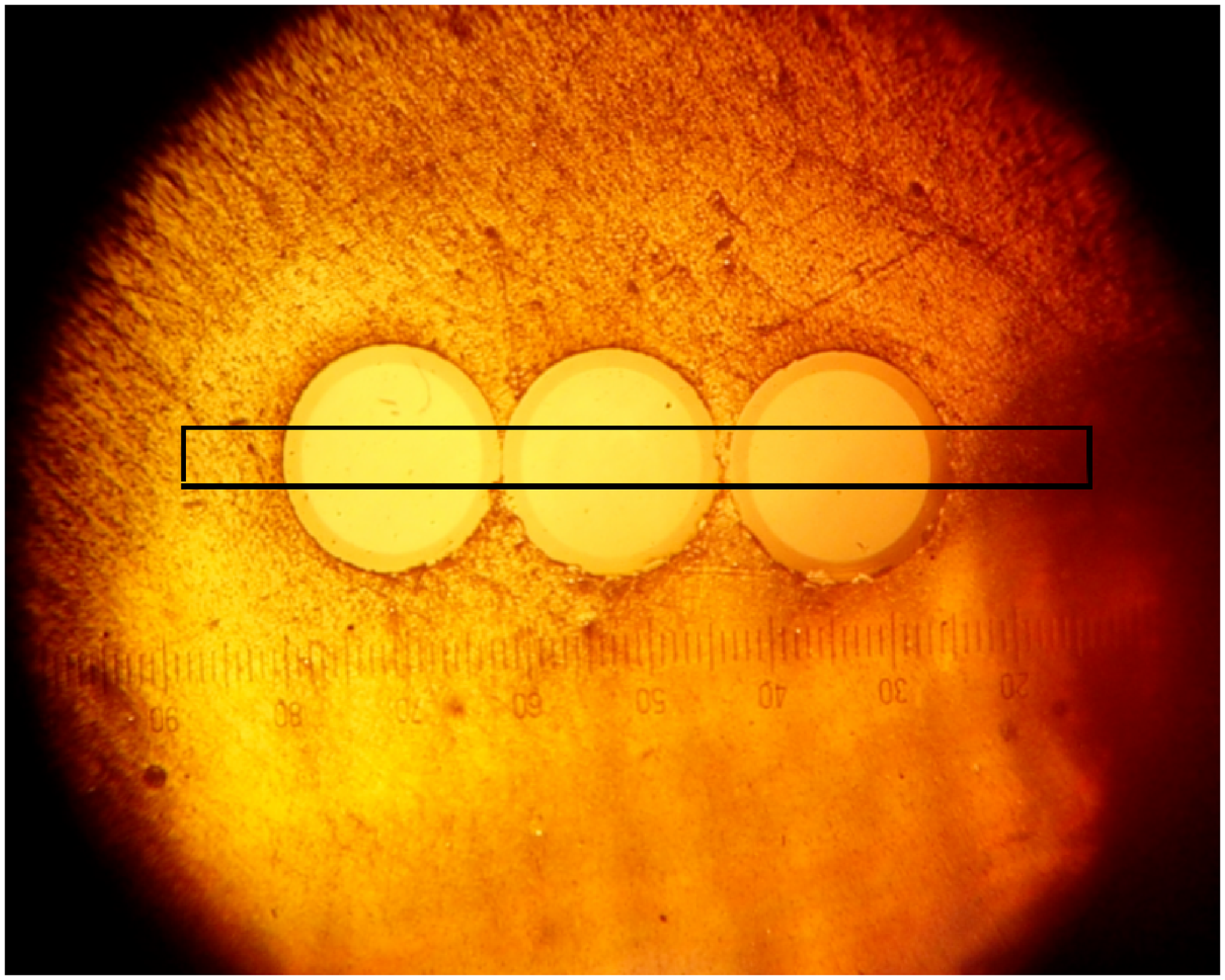}
\caption{Photographs of the 2006 fiber slicer object input (left), fiber slicer pseudo slit with calibration fibers on each end (middle) for 2006 August data, and the fiber arrangement for 2007 November data with object, blank, and calibration fiber (right).  The box in the 2007 November fiber arrangement shows the approximate location of a 100-\mm slit.} \label{plotthree}
\end{center}
\end{figure}

For the 2006 August dataset a fiber slicer was used for the object input.  Our slicer uses 100-\mm\ core, 125-\mm\ cladding diameter fibers; this yields a geometrical throughput of~49\% after the Acrylate buffers are removed from the constituent fibers. The left panel of Figure~\ref{plotthree} displays the input of the seven slicer fibers, and the middle panel is the output of the slicer fibers formatted into a psuedo-slit. The two outermost fibers on each end of the slit are the calibration lamp feeds. A single fiber spacer is used to separate each calibration fiber from the seven object fibers. Initial measurements of the laser output indicated acceptable focal ratio degradation.  In October 2007, we implemented a configuration with 300~\mm\ core fibers.  The object fiber is separated from the calibration fiber by a single fiber spacer (see right-hand image in Fig.~\ref{plotthree}).  To achieve our desired resolution the output end of the new input fiber cable was butted up against a 100~\mm\ slit on a 6-mm diameter stainless steel plate.
In both instrument setups, we vibrated the fibers to minimize the effects of modal noise (Baudrand \& Walker, 2001).  By experiment, we found that this reduced the modal noise below the level of concern for this work.  We have conducted extensive experiments on eliminating modal noise in the NIR, which will be detailed in a subsequent paper.

\subsection{Control System}

The \'{e}chelle and cross-dispersor gratings are mounted on rotational stages and the camera is mounted on a linear stage.  All three stages are driven with a DC servo motor with incremental encoder position  feedback.  A Newport ESP 300 motion controller with a RS232 interface to a PC are used for setting and tracking  positions. Although we would prefer not to change the positions of the optics, the small size of the detector compared to the area covered by the spectral range of interest requires precision control of both the \'{e}chelle and cross disperser.  The ability to establish and adjust focus with a precision control system is also highly desirable, although in practice the focus rarely needs adjustment.

\subsection{Calibrations}

The calibration unit contains Th-Ar, Ar, Kr, Ne and Xe pen lamps as well as a Quartz Iodine flat field lamp. Optical fibers are employed to combine the Nobel-gas lamp signals into a single integrating fiber. Th-Ar is the calibration of choice for most high resolution spectrographs, but combining the Ar, Kr, Ne and Xe lamps increased the number of available reference lines in the longer end of Pathfinder's waveband compared to the Th-Ar calibration. For the Y-band wavelength range presented here, we used the Ar pen lamp (2006) and Th-Ar lamp (2007).

\subsection{Solar Feed}

The signal required for our experiment is the integrated solar disk; this is obtained by imaging the Sun onto a 50-m long Polymicro FIP, 300~\mm\ core fiber which scrambles spatial information. The input end of the cable has a 2.5~mm ball lens with a 10$^{\circ}$ field of view, so tracking errors should not compromise obtaining integrated solar disk light.  Unfortunately, we found that the field of view was so large that light scattered by nearby clouds noticeably degraded the quality of our signal and so observations were only attempted when skies were mostly clear.  Clouds that do not uniformly cover the Sun during an integration can lead to a significant solar rotation residual in the data as solar rotation is about 100 times greater than what we are trying to measure.

\section{Data Reduction \& Analysis}

\begin{figure}
\begin{center}
\includegraphics[height=165mm,angle=-90]{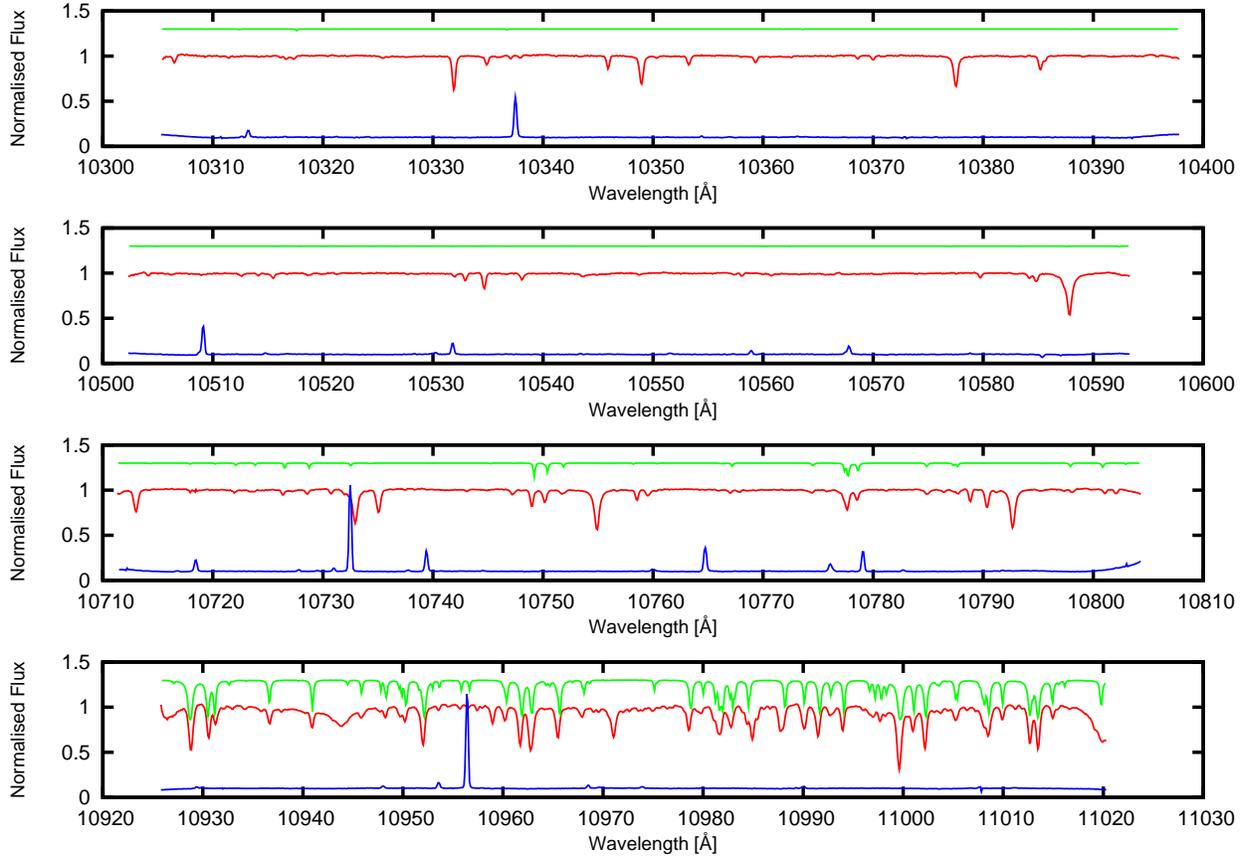}
\end{center}
\caption{Spectra and wavelength coverage of orders \novapmin (bottom) through \novapmax (top) of the 2007 November 2 data set. Within each wavelength region we show the telluric model spectrum (top), observed (solar and telluric) spectrum (middle) and the Th-Ar emission reference spectrum divided by a factor of 10 (bottom).}
\label{plotfour}
\end{figure}

To explore the issues involved in precision radial velocities in the NIR, we used the Earth's rotation signal, found in integrated Sunlight.  This signal has an amplitude of several hundred \ms\thinspace and can be measured on any clear day in the laboratory avoiding the added time and expense of going to a remote telescope. Our program concentrated on a region of the Y band selected to have (1) relatively few strong telluric features, (2) numerous solar features, and (3) enough calibration lines to track spectrograph instabilities. The spectral region used is illustrated in Figure~\ref{plotfour}, and provided wavelength coverage between 1.03~\mm~and 1.103~\mm.

\begin{figure}
\begin{center}
\includegraphics[width=165mm,angle=-0]{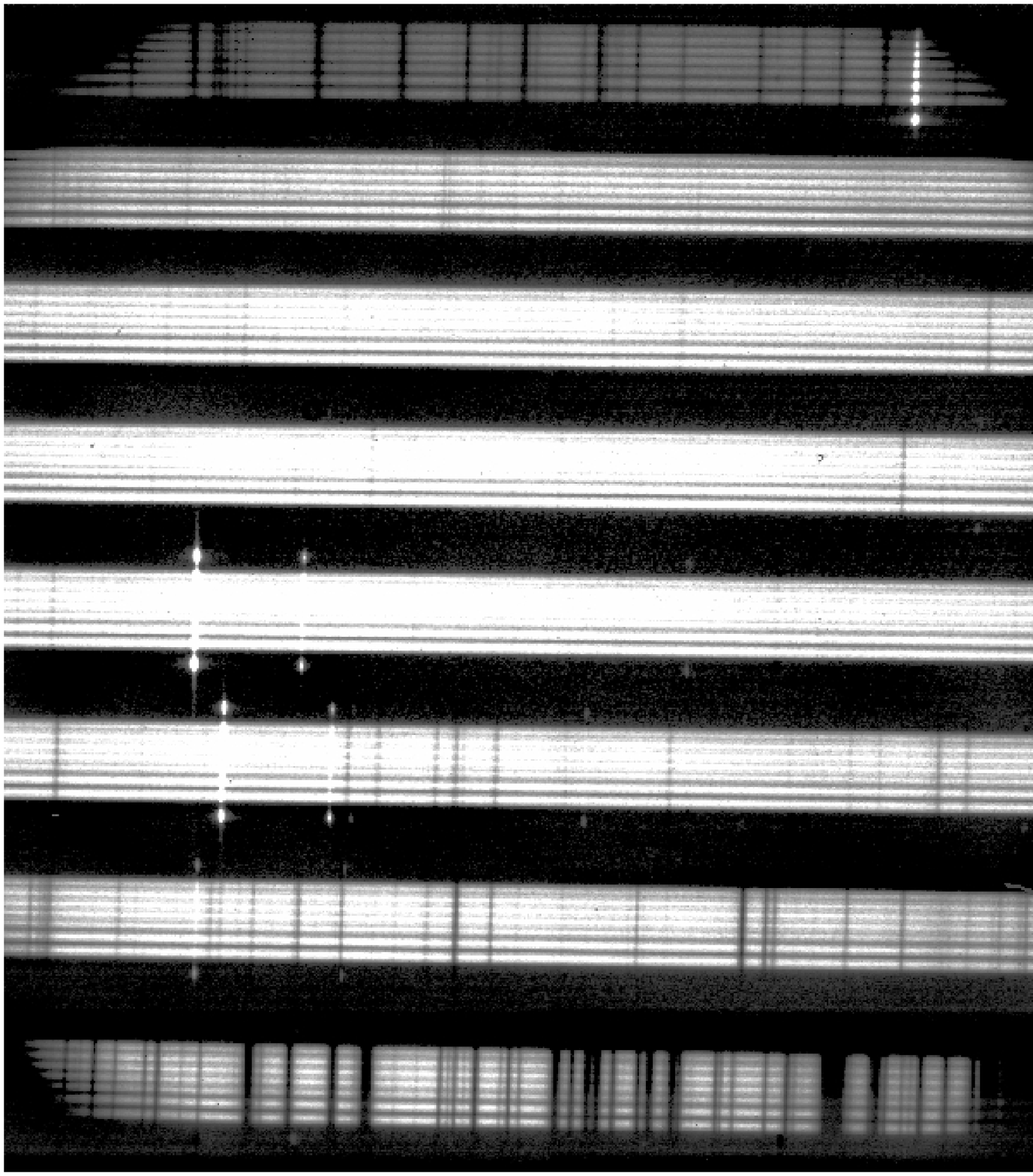}
\end{center}
\caption{Y-band spectrum showing orders 49 (bottom) to 56 (top) recorded by Pathfinder (left - 2006 August; right - 2007 November). Wavelengths increase from right to left and from top to bottom. The orders used for the analysis presented are orders \augapmin \& \augapmax (2006 August) and orders \novapmin-\novapmax (2007 November).}
\label{plotfive}
\end{figure}

While we have experimented with numerous instrument configurations, we will only discuss initial and most recent tests.  On 2006 August 16, we acquired $781$ $10$-second exposures using the fiber setup shown in the left and middle images of Figure~\ref{plotthree} (hereafter, the 2006 data set).  This slicer produced spectra with overlapping solar and calibration feeds, flanked by an additional calibration feed, as seen in the left-hand image of Figure~\ref{plotfive}.  On 2007 November 2, we took three sets of $300$ successive 10-second exposures with solar and calibration spectra in separate fibers (hereafter, the 2007 data set), as can be seen in the right-hand images of Figs.~\ref{plotthree} and~\ref{plotfive}.  In acquiring both data sets, the first five frames of each sequence were discarded to minimize the effect of persistence due to the usually saturated image that occurs when the detector is not read out for some time.

We made four significant alterations to the instrument setup between these two observations.  We (1) altered the fiber setup (as mentioned) to improve S/N in the Th-Ar lines, (2) improved the thermal insulation around the spectrograph, to minimize temperature-dependent instrumental drifts, (3) adjusted the wavelength region to minimize the number of telluric features, and (4) swapped a 1300-\mm~filter for an 1800-\mm~filter.  The last of these modifications enabled us to test suppression of modal noise in the H-band by vibration, but significantly increased the thermal background.

\begin{figure}
\begin{center}
\includegraphics[height=165mm,angle=-90]{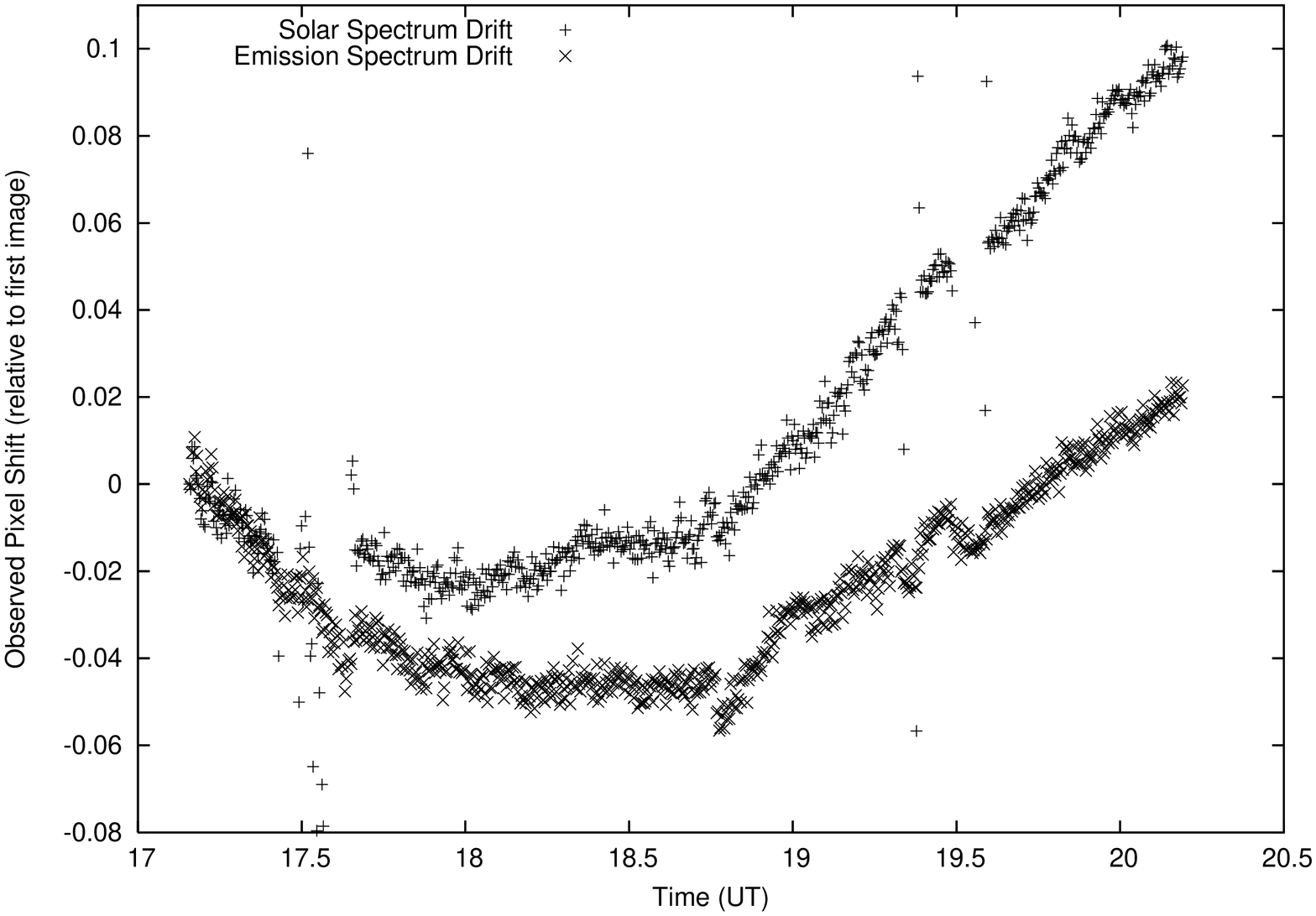}
\end{center}
\caption{Instrumental and solar drift from cross-correlation of order \augapmin of the 2006 August solar and emission line spectra relative to the first observation. In this figure, a drift of 0.1 pixels corresponds to a change in velocity of 241\ms.  The dominant shape of both curves is due to instrumental drift while the difference in magnitude of the solar drift when compared with the instrumental drift is the result of the Earth's rotation.  Note the breaks in the solar drift caused by the presence of clouds.}
\label{plotsix}
\end{figure}

While absolute wavelength calibration against a template Th-Ar or Iodine absorption is essential for any long-term program, we were interested in looking at changes within a clear observing run. No attempt was made to correct the small deparatures from linearity present in the raw spectra and establish an absolute wavelength scale in the 2006 data set.  Using the calibration lines present and the solar lines, we obtain a scale of $\sim 2.4$ \kms per pixel. Figure~\ref{plotsix} shows the amplitude of drift in reference lamp lines to be $\sim 300$~\ms\thinspace or 0.12 pixels. The corresponding drift in the solar lines is also shown and can be seen to be of greater magnitude. By subtracting simultaneous pairs of solar-reference points, we account for drifts in the spectrograph, leaving only the radial velocity variation of the Sun due to the Earth's rotation.  We ignore the acceleration due to the Earth's elliptical orbit, as well as that due to the apparent motion of the Sun at our latitude due to the obliquity of the ecliptic, as both are small during the observation period.

With the 2007 data set, the same \'{e}chelle orders were used, but shifted to slightly bluer wavelengths, to increase the number of observed Th-Ar reference lines. The Th-Ar lamp current was also increased to augment the weaker lines. Nevertheless, the few lines and incomplete indentification of Th and Ar emission lines in this region of the electromagnetic spectrum only enabled a wavelength calibration using several lines to be made. From Th-Ar and telluric wavelength calibrations, we estimate a pixel resolution of $2.5$ \kms for the 2007 instrumental configuration.

Both data sets were reduced independently, once with Starlink\footnote{http://starlink.jach.hawaii.edu/} routines, and again with {\sc iraf}\footnote{http://iraf.noao.edu/} routines, for the purpose of comparing the quality of the reduction routines and help eliminate any systematics that might be introduced in the reduction process.  All data were also analyzed using two different cross-correlation routines --- {\sc hcross} (using the algorithm developed by Heavens, 1993) in Starlink, and {\sc fxcor} (Fitzpatrick, M.J., 1993) in {\sc iraf}.  The similarity of the results gave us confidence that both packages are viable at this level and our results are sound.  In each program, we analyzed regions that were free of obvious telluric contamination, and cross-correlated the individual spectra relative to the first exposure.  In the 2006 data set, orders \augapmin \& \augapmax were used, while in the 2007 data set, orders \novapmin-\novapmax were used.  The 2006 \& 2007 data sets had $13$ and $31$ solar lines with absorption features greater than 5\% of the continuum, respectively, and $2$ and $16$ emission lines with more than $1500$ counts above background, respectively.

\subsection{Results \& Discussion}

\begin{figure}
\begin{center}
\includegraphics[height=165mm,angle=-90]{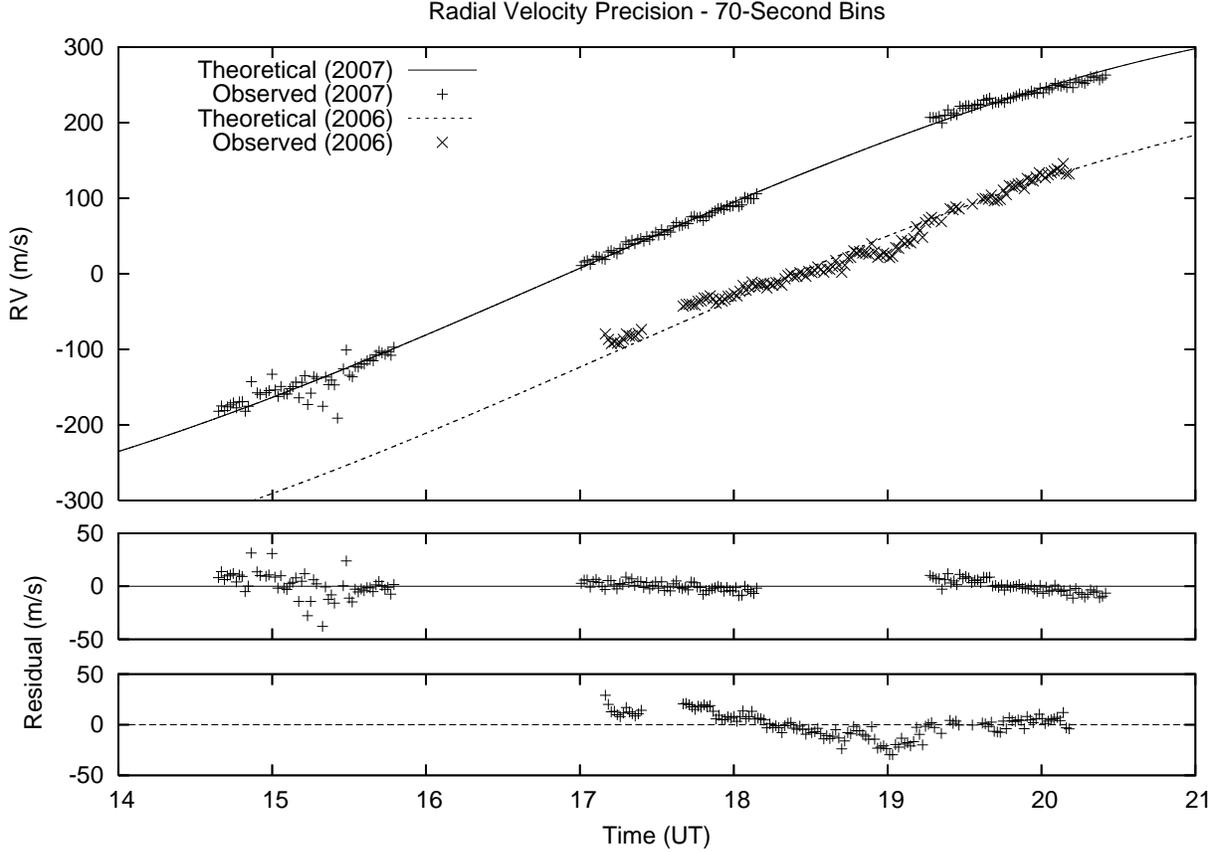}
\end{center}
\caption{Top panel: The radial velocity of the Sun due to the Earth's rotation on 2007 November 2 (top curve) and 2006 August 16 (bottom curve), binned in groups of 5 images. The points are the cross-correlation measurements corrected for instrumental drift which amounted to $\sim 200$\ \ms\thinspace (0.083 pixels) over the three hours of observations for the 2006 data set and $\sim 350$~\ms\thinspace (0.140 pixels) over six hours for the 2007 data set. The measurements have been arbitrarily shifted by small amounts (since we have not performed absolute radial velocity measurements) to best match the theoretical curve. The curves show the predicted sinusoidal variation with respect to apparent solar time. The bottom panels show the residuals for the two data sets.}
\label{plotseven}
\end{figure}

\begin{figure}
\begin{center}
\includegraphics[height=165mm,angle=-90]{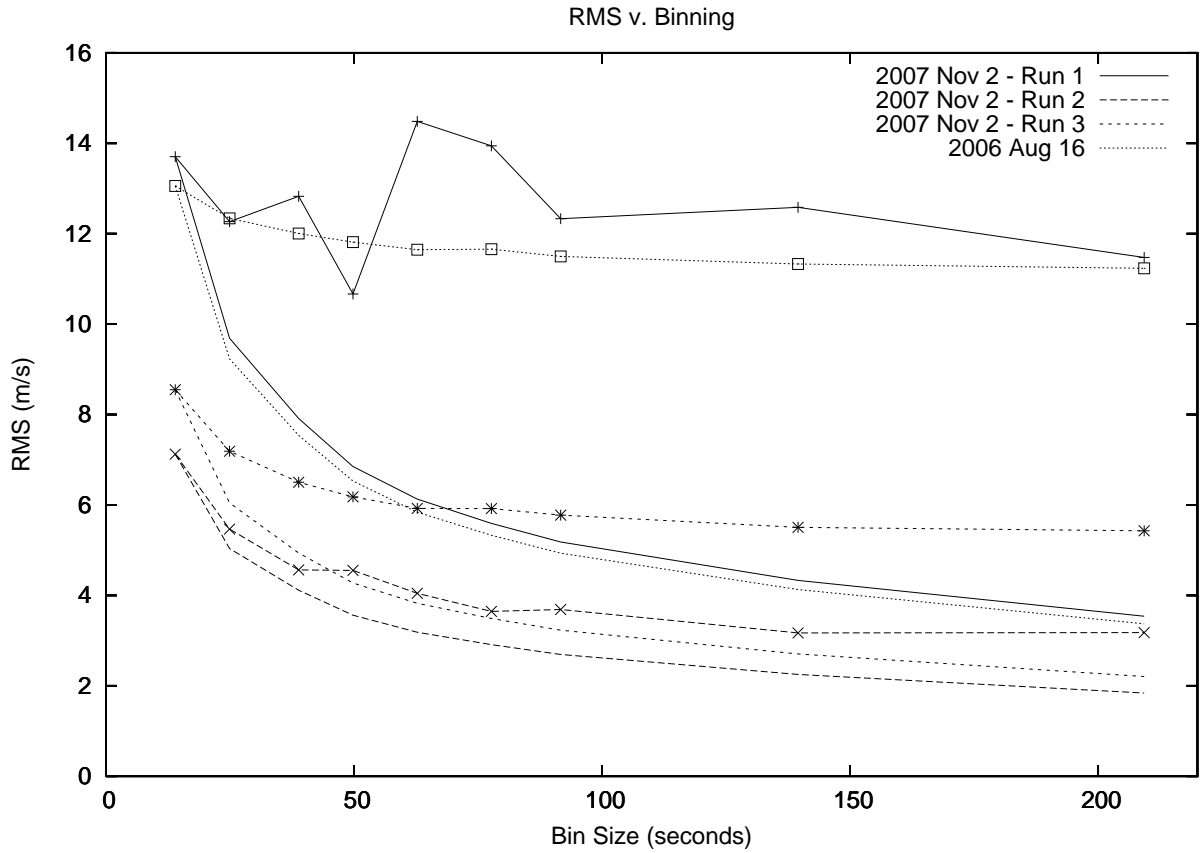}
\end{center}
\caption{RMS values in \ms\thinspace for the three observational sequences of the 2006 \& 2007 data sets (with data points) with various amounts of binning, along with the theoretical $\sqrt{N}$ (no data points) drop in the measured RMS for each run.  While the exposures are all 10-s each, read-out time increases the time between successive images to 14 s.}
\label{ploteight}
\end{figure}

Our results are presented in Figure~\ref{plotseven}, where we have plotted the theoretical and observed RV curves in the top panel (providing a $100$~\ms\thinspace offset to the 2006 data set for clarity), and the residuals of the 2006 and 2007 data sets in the middle and bottom panels, respectively.  Gaps in the 2006 data set are due to periods of cloud cover, while gaps in the 2007 data set were times during which we took various calibration (flatfield, dark, and wavelength) data.  We achieved an overall RMS of $13.1$~\ms\thinspace for the 2006 data set.  We ignored those intervals where the average pixel counts dropped more than $50\%$ below the mean (\emph{i.e.}, periods of cloud coverage).  For the 2007 data set we attain an RMS of $7.1$~\ms\thinspace for the second run using {\sc fxcor} ($9.5$~\ms\thinspace using {\sc hcross}).  Binning the 2007 data in sets of $5$ images improves our results, reducing the RMS to $4.1$~\ms\thinspace for the second run, as can be seen in Figure~\ref{ploteight}.  However, binning in sets of $N$ images does not improve our results as $\sqrt{N}$, as would be expected if random error dominated.  As we note below, there is a systematic error present that puts a floor on the precision gain from binning.  For comparison, $\sqrt{N}$ plots can be seen in Fig.~\ref{ploteight}.  These 1-minute RMS values are an order of magnitude above those obtained in the long-running {\sc bison} ($0.2$~\ms) and {\sc golf} ($0.6$~\ms) projects (based on table 1 of Butler et al., 2004).  The additional noise in the first run of the 2007 November data is due to the presence of a few scattered clouds, which caused the average pixel counts to vary up to $50\%$.  The second and third run had less than a $1\%$ variation.  The third run (with an RMS of $8.5$~\ms) shows a distinct linear trend which also appears (to a lesser extent) in run two.  The cause of this trend is not yet understood but clearly something that is monotonic with time and not compensated by the calibration scheme.  We are upgrading our system to include precision temperature monitoring at several points and, as discussed below, we are also updating our calibration system to aid in minimizing this type of error.

There are a number of reasons to believe that considerable improvements are possible. Our 2007 analysis is based on relatively modest S/N data (for the Sun) and a small wavelength interval.  Also the Sun has few and relatively weak infrared features. Identical measurements on an M dwarf would yield many more absorption lines, giving rise to a much stronger cross-correlation signature. The use of a larger infrared array and an optimized cross-disperser would allow simultaneous coverage of the Y \& J bands increasing the spectral coverage by approximately a factor of 10 resulting in  increased prescion.

The greatest improvement will come from using hallow cathode lamps with a larger infrared line density.  To this end, we explored the Uranium spectra, which revealed $97$ Th-Ar lines and $226$ U lines between 839.5 to 897.5 nm (Palmer et al., 1980).  This encouraged us to procure a UAr hollow cathode lamp; a preliminary observation in the Y-band revealed a similar two-to-one ratio of U to Th lines.  Using both lamps simulanteously in future experiments will undoubtedly improve our measured precision.

\section{Conclusion}

For the first time in the NIR, we have demonstrated RMS radial velocity precisions below $10$ \ms\thinspace over a period of several hours, using $10$-s observations of the Sun. There are a number of improvements that we hope to implement when additional funding is secured.  Foremost is to install a R4 \'{e}chelle appropriately sized and a VPH or standard cross-disperser grating optimized for the $1-1.3$~\mm region.  These two items in themselves will lead to close to an order of magnitude increase in throughput.   In addition, we will update our calibration system to allow us to move from precision acceleration measurements over a short period of time to precise velocity measurement over weeks to months.    The upgraded calibration system will include both Th-Ar and U-Ar hollow cathode lamps as well as a gas cell that is illuminated by a Q-I lamp.  While we do not intend to use a gas cell simultaneously in line with the object observations as pioneered by Butler et al (1996) in the visible, it will allow a better absolute velocity determinations as well as a means to track systematics in the hollow cathode lamps.   We are implementing the capability to rapidly switch between the target spectrum and the gas cell to allow calibrations between target spectra.  We will initially test HF and water vapor gas cells in the 1.0 to 1.25 micron range but note that no single gas has a large number of deep lines over the desired wavelength range.  Adding the U-Ar hollow cathode lamp will increase the number of lines at least a factor of two and allow less dependence on Ar lines.  We also will investigate the effects persistence may have on the achievable precision.  With the noted improvements we will be ready to move this instrument to the Hobby-Eberly Telescope for tests on M dwarfs to firmly establish the viability of NIR precision radial velocity spectroscopy in these objects.

\section{Acknowledgements}

The effort through September 2006 was partially supported by a Gemini Design Study for the Precision Radial Velocity Spectrometer and National Science Foundation Grant AST06-07634. We acknowledge many valuable conversations with other members of the PRVS team, in particular, John Rayner, Bill Dent, Adrian Webster and Chris Tinney.  We also acknowledge the continuing encouragement by Gemini.  HRA Jones and J. Barnes are supported by PPARC grant PP/D0000920/1.  Finally, we thank Don Schneider and NSF Grant AST03-07582 for their support.

\section{References}

\noindent
Baudrand, J., Walker, G.A.H., 2001, PASP, 113, 851\\
Becker, J.M., 1976, Nature, 260, 227\\
Blake, C. H., Charbonneau D., White R.J., Marley M.S., Saumon D., 2007,
ApJ, 666, 1918 \\
Butler, R. P., Marcy, G. W., Williams, E., McCarthy, C., Dosanjh, P., Vogt, S. S., 1996, PASP, 108, 500\\
Butler, R.P. et al., 2004, ApJL, 600, 75 \\
Campbell~B., Walker~G.A.H., 1979, PASP, 91, 540\\
Campbell B., Walker G.A.H., Yang S., 1988, ApJ, 331, 902\\
Deckker, H. et al, 2000, Proc.SPIE, 4008, 534\\
Fitzpatrick, M.J., 1993, ASPC, 52, 472 \\
Gray, D.F., Brown, K.I.T., 2006, PASP, 118, 399\\
Heavens, A.F., 1993, MNRAS, 263, 735 \\
Hinkle, K.H. et al. 2003, Proc.SPIE, 4834, 353\\
Johnston, K. 2006, MSci Project, University of St. Andrews\\
Kasting, J.F., Catling, D., 2003, ARA\&A, 41, 429 \\
Koch~A., Woehl~H., 1984, A\& A, 134, 134\\
Martin E.L. et al, 2006, ApJ, 644, 75 \\
McLean, I.S., et al, 1998, SPIE, 3354, 566 \\
Palmer, B.A., Keller, R.A., Engleman, R., Los Alamos Informal Report LA-8251-MS, 1980 \\
Pepe, F. et al, 2000, Proc.SPIE, 4008, 582\\
Schroeder, D.J., 2000, Astronomical Optics (San Diego: Academic Press)\\
Stassun, K.J., Mathieu, R.D., \& Valenti, J.A., 2006, Nat, 440, 311 \\
Tinney, C.G. et al., 2001, ApJ, 551, 507\\
Tull, R.G. 1998, Proc.SPIE, 3355, 387\\
Vogt, S.S. et al, 1994, Proc.SPIE,  2198, 362\\
Wolszczan, A., Frail, D.A., 1992, Nature, 355, 145\\
\end{document}